\documentclass[aps,prl,twocolumn,superscriptaddress,nofootinbib]{revtex4-2}

\usepackage{graphicx}
\usepackage{dcolumn}
\usepackage{bm}
\usepackage{amsmath}
\usepackage{amsfonts}
\usepackage{amssymb}
\usepackage{physics}
\usepackage{float}
\usepackage[citecolor=blue,linkcolor=blue,urlcolor=blue]{hyperref}
\usepackage[ocgcolorlinks]{ocgx2}
\usepackage{placeins}
\usepackage{multirow}
\usepackage{dcolumn}

\newcommand{\be}{\begin{equation}}
\newcommand{\ee}{\end{equation}}

\newcommand{\fmiq}{\, \text{fm}^{-3}}

\newcommand{\mev}{\, \text{MeV}}
\newcommand{\prlsection}[1]{\emph{#1.}--}
\newcommand{\rn}{{\rm n}}
\newcommand{\rp}{{\rm p}}

\begin{document}

\title{Neutron star matter as a dilute solution of protons in neutrons}

\author{J. Keller}
\email{j.keller@theorie.ikp.physik.tu-darmstadt.de}
\affiliation{Technische Universit\"at Darmstadt, Department of Physics, D-64289 Darmstadt, Germany} 
\affiliation{ExtreMe Matter Institute EMMI, GSI Helmholtzzentrum f\"ur Schwerionenforschung GmbH, D-64291 Darmstadt, Germany}

\author{K. Hebeler}
\email{kai.hebeler@physik.tu-darmstadt.de}
\affiliation{Technische Universit\"at Darmstadt, Department of Physics, D-64289 Darmstadt, Germany} 
\affiliation{ExtreMe Matter Institute EMMI, GSI Helmholtzzentrum f\"ur Schwerionenforschung GmbH, D-64291 Darmstadt, Germany}
\affiliation{Max-Planck-Institut f\"ur Kernphysik, Saupfercheckweg 1, D-69117 Heidelberg, Germany}

\author{C. J. Pethick}
\email{pethick@nbi.dk}
\affiliation{The Niels Bohr International Academy, The Niels Bohr Institute, University of Copenhagen, Blegdamsvej 17, DK-2100 Copenhagen \O, Denmark}
\affiliation{NORDITA, KTH Royal Institute of Technology and Stockholm University, Hannes Alfv\'ens v\"ag 12, SE-106 91 Stockholm, Sweden}

\author{A. Schwenk}
\email{schwenk@physik.tu-darmstadt.de}
\affiliation{Technische Universit\"at Darmstadt, Department of Physics, D-64289 Darmstadt, Germany} 
\affiliation{ExtreMe Matter Institute EMMI, GSI Helmholtzzentrum f\"ur Schwerionenforschung GmbH, D-64291 Darmstadt, Germany}
\affiliation{Max-Planck-Institut f\"ur Kernphysik, Saupfercheckweg 1, D-69117 Heidelberg, Germany}

\begin{abstract}
Neutron stars contain neutron-rich matter with around $5\%$ protons at nuclear saturation density. In this Letter, we consider equilibrium between bulk phases of matter based on asymmetric nuclear matter calculations using chiral effective field theory interactions rather than, as has been done in the past, by interpolation between the properties of symmetric nuclear matter and pure neutron matter. Neutron drip (coexistence of nuclear matter with pure neutrons) is well established, but from earlier work it is unclear whether proton drip (equilibrium between two phases, both of which contain protons and neutrons) is possible. We find that proton drip is a robust prediction of any physically reasonable equation of state, but that it occurs over a limited region of densities and proton fractions. An analytical model based on expanding the energy in powers of the proton density, rather than the neutron excess, is able to account for these features of the phase diagram.
\end{abstract}

\maketitle

\prlsection{Introduction} In neutron stars, matter is extremely neutron rich. Around nuclear saturation density, $n_0 = 0.16\fmiq$, in the inner crust and outer core, it consists primarily of neutrons, protons, and electrons, with typical proton fractions $\sim 5\%$~\cite{Haensel2007}. A detailed understanding of the equation of state (EOS) and of the phases of matter under such conditions is crucial for determining the properties of the neutron star crust, among them whether or not the so-called pasta phases with string-like and plate-like nuclei can exist near the inner boundary of the crust.

The typical approach to calculating the properties of such matter has been to make simple interpolations between those of symmetric nuclear matter and those of pure neutron matter. However, studies of dilute Fermi gases show that the dependence of the energy density on proton density $n_\rp$ for small proton concentrations is more complicated with, e.g., terms varying as $n_\rp^{7/3} \ln n_\rp$~\cite{Kanno1970}. Recently, chiral effective field theory (EFT) calculations of asymmetric nuclear matter with low proton concentrations have been performed~\cite{Keller2023}, which avoid such simplistic interpolations. In this Letter, we combine these with analytical considerations to determine the phase diagram of low-density neutron star matter.

Indeed, one point on which there has been disagreement over the years is the question of whether or not two bulk phases, both of which contain protons as well as neutrons, can coexist, so-called proton drip.  In Ref.~\cite{BBP} no evidence for proton drip was found, while in other works it did occur~\cite{Pethick1995,Douchin2000}. We will show that proton drip is a universal feature of any realistic EOS, but that it occurs over limited ranges of densities and proton fractions.

\prlsection{Chiral EFT calculations} We investigate the phase structure of neutron-rich matter at zero temperature based on our microscopic calculations of asymmetric nuclear matter from chiral EFT interactions~\cite{Keller2023}. Using many-body perturbation theory, we calculate the energy density $\varepsilon = E/V = \langle H  \rangle/V$ of the ground state of spatially uniform matter, as a function of the neutron density $n_\rn$ and the proton density $n_\rp$. The Hamiltonian $H = T + V_{\rm NN} + V_{\rm 3N}$ includes the kinetic energy $T$, two-nucleon (NN) interactions $V_{\rm NN}$, and three-nucleon (3N) interactions $V_{\rm 3N}$. We include all chiral EFT interactions up to next-to-next-to-next-to-leading order (N$^3$LO) with the NN potentials from Ref.~\cite{EMN2017} and 3N interactions fit to the $^3$H binding energy and the empirical saturation region in Ref.~\cite{Drischler2019}. Our main results are based on N$^3$LO NN and 3N interactions with a cutoff $\Lambda = 450 \mev$, but we also consider results at N$^2$LO to test the sensitivity to the chiral EFT truncation.

The nuclear matter calculations include contributions to the energy up to third order in many-body perturbation theory around a Hartree-Fock reference state (for details see Ref.~\cite{Keller2023}). In addition, we use a Gaussian process (GP) emulator for the energy~\cite{Keller2023}. The GP allows the evaluation of the EOS and derivatives of it for arbitrary conditions within the calculated range without multi-dimensional interpolation. The pressure is then given by $P=\left.n^2 \partial (\varepsilon/n)/\partial n\right|_x$, where $n=n_\rn + n_\rp$ is the baryon density and $x=n_\rp/n$ the proton fraction, and the neutron and proton chemical potentials by $\mu_\rn=\left.\partial \varepsilon/\partial n_\rn\right|_{n_\rp}$ and $\mu_\rp=\left.\partial \varepsilon/\partial n_\rp\right|_{n_\rn}$, respectively.

\prlsection{Coexistence} Since Coulomb and surface energies are generally small compared with bulk energies, we focus on equilibrium between bulk phases. For coexistence of two phases, denoted by 1 and 2 with nucleon densities $n_\rn^{(1)}, n_\rp^{(1)}$ and $n_\rn^{(2)},n_\rp^{(2)}$, the pressures and chemical potentials must satisfy the conditions
\begin{align}
P^{(1)} &= P^{(2)} \,, \label{eq:pressureequm} \\
\mu_\rn^{(1)}  &\ge \mu_\rn^{(2)} \,, \label{eq:mu_nequm} \\
{\rm and} \quad \mu_\rp^{(1)} &\ge \mu_\rp^{(2)}  \,. \label{eq:mu_pequm}
\end{align}
Here we take phase 2 to be the higher-density phase that contains both neutrons and protons, so that $P^{(2)}$, $\mu_\rn^{(2)}$, and $\mu_\rp^{(2)}$ depend on both neutron and proton densities. We first consider the case in which phase 1 contains only neutrons (neutron drip), and then Eq.~(\ref{eq:mu_nequm}) is an equality. For nuclear matter to be in equilibrium with a pure neutron phase (with $n_\rp^{(1)} = 0$), in addition to equal neutron chemical potentials, it is necessary that $\mu_\rp^{(1)} > \mu_\rp^{(2)}$, otherwise it is energetically favorable for protons to start populating the initially pure neutron phase until Eq.~(\ref{eq:mu_pequm}) becomes an equality. In the case that also the proton chemical potentials are equal, the proton density in phase~1 is finite, $n_\rp^{(1)} > 0$, and phase 1 consists of dripped protons in addition to dripped neutrons.

\begin{figure}[t]
  \centering
  \includegraphics[clip=,width=\columnwidth]{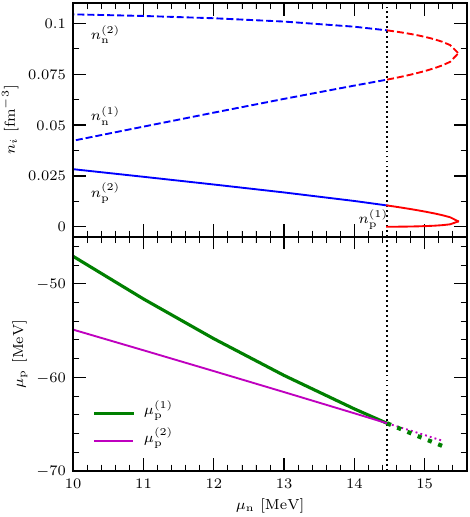}
  \caption{Upper panel: Coexistence of neutron matter with nuclear matter (neutron drip, blue lines) and neutron and proton matter with nuclear matter (proton drip, red lines) as a function of the neutron chemical potential. The neutron and proton densities in the two phases are shown as dashed and solid lines, respectively. Lower panel: Proton chemical potentials for coexistence of neutron matter with nuclear matter. The vertical dotted line shows the neutron chemical potential where both proton chemical potentials are equal, so that proton drip occurs for higher neutron chemical potentials. All results are for the N$^3$LO asymmetric matter EOS.}
  \label{fig:composition-and-chemical-potentials}
\end{figure}

The solutions to Eqs.~(\ref{eq:pressureequm})--(\ref{eq:mu_pequm}) for the N$^3$LO asymmetric matter EOS are shown in Fig.~\ref{fig:composition-and-chemical-potentials} as a function of the neutron chemical potential. The densities for the neutron drip phase, $n_\rn^{(1)}$, $n_\rn^{(2)}$, and $n_\rp^{(2)}$, are shown as blue lines, while the densities for proton drip (with additional $n_\rp^{(1)} > 0$) are shown in red when coexistence is possible.
The lower panel shows the proton chemical potentials for the two phases. For $\mu_\rn \gtrsim 14.47 \mev$ the proton chemical potential in neutron matter $\mu_\rp^{(1)} = \mu_\rp(\mu_\rn, n_\rp = 0)$ is smaller than in nuclear matter 
\be
\mu_\rp^{(1)} < \mu_\rp^{(2)} = \mu_\rp(\mu_\rn, n_\rp^{(2)})\,,
\ee
so that it is energetically favored for protons to move from nuclear matter to neutron matter (proton drip). In this region, low-density neutron and proton matter coexists with high-density nuclear matter. With increasing neutron chemical potential (increasing total density), the low- and high-density solutions merge in the top panel of Fig.~\ref{fig:composition-and-chemical-potentials}. This is where the inhomogeneous proton drip phase ends and matter becomes uniform.

\begin{figure}[t]
  \centering
  \includegraphics[clip=,width=\columnwidth]{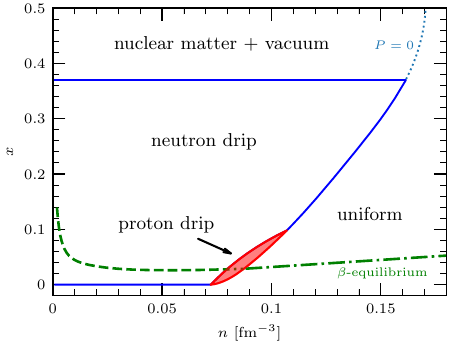}
  \caption{Phase diagram as a function of the total density $n$ and the proton fraction $x$ at N$^3$LO. The neutron drip and proton drip phases are given by the regions encompassed by the blue and red lines. At high densities, matter is in the uniform phase, and for proton fractions $x \gtrsim 0.37$, nuclear matter is self-bound (for densities below the zero pressure, $P=0$, dotted line) but does not fill the entire volume (nuclear matter + vacuum). In addition, we show the composition of matter in beta-equilibrium (green dot-dashed line in the uniform phase, and green dashed and dotted lines in the neutron and proton drip phases, respectively).}
  \label{fig:phase-diagram}
\end{figure}

\prlsection{Phase diagram} The coexistence of dripped neutrons and protons with nuclear matter occurs along lines in the $(n_\rn, n_\rp)$ plane with
\begin{align}
n_\rn &= n_\rn^{(1)} (1 - u) + n_\rn^{(2)} u \,, \label{volfrac1} \\
n_\rp &= n_\rp^{(1)} (1 - u) + n_\rp^{(2)} u \,, \label{volfrac2}
\end{align}
where $u \in [0, 1]$ is the volume fraction (and $n_\rp^{(1)} = 0$ for neutron drip).
The corresponding phase diagram in the $(x, n)$ plane is shown in Fig.~\ref{fig:phase-diagram}.
As expected, neutron drip (enclosed by the blue line) is possible for a large region in proton fraction and density.
The end of the neutron drip region at $x\approx 0.37$ is reached for $\mu_\rn = 0$. For larger proton fractions and densities below the zero pressure, $P = 0$, dotted line, nuclear matter is self-bound but does not fill the entire volume. In addition to these established features of the phase diagram, we find a region at lower proton concentration where proton drip is possible. This is shown by the red-shaded region in Fig.~\ref{fig:phase-diagram}. We have also checked that the spinodal line, which marks the boundary of the region where matter is unstable to density fluctuations, is contained within the neutron and proton drip regions.

In order to explore whether proton drip is relevant to neutron stars, we must include the effect of electrons, which we shall treat as a uniform background of negative charge. The condition for beta equilibrium in the homogeneous, electrically neutral phase is  $\mu_\rn\left(x(n), n\right) = \mu_\rp\left(x(n), n\right) + \mu_{\text{e}}\left(x(n), n\right)$. The density of electrons $n_\text{e}$ is equal to $n_\rp$ and their chemical potential is given by $\mu_{\text{e}} \approx (3\pi^2 n_{\text{p}})^{1/3}$ since they are ultrarelativistic. This is shown by the green dot-dashed line in Fig.~\ref{fig:phase-diagram}. As the density decreases, the beta equilibrium line enters the proton drip phase, so that proton drip is relevant for neutron stars. We also show in Fig.~\ref{fig:phase-diagram} the composition in beta equilibrium in the proton drip and neutron drip phases. Finally, we have checked that the neutron and proton drip phases have the lowest energy for a given density $n$.

\begin{figure}
  \centering
  \includegraphics[clip=,width=\columnwidth]{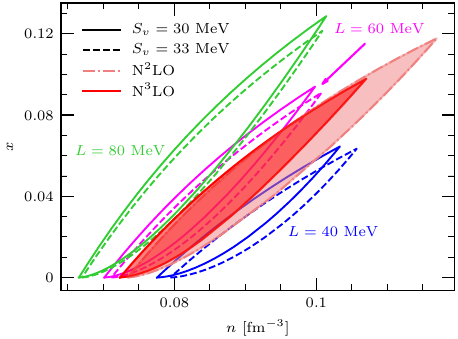}
  \caption{Comparison of the proton drip region at N$^3$LO (as shown in Fig.~\ref{fig:phase-diagram}) to results at lower order, N$^2$LO (light red region encompassed by the dash-dotted line), as well as from a phenomenological parametrization of the energy~\cite{Hebeler2013} using reasonable ranges of the symmetry energy $S_v$ (solid and dashed lines) and the $L$ parameter (different colors).}
  \label{fig:robustness}
\end{figure}

\prlsection{Robustness} Next we explore how the phase diagram depends on the EOS. In Fig.~\ref{fig:robustness} we enlarge the proton drip region and compare the region for the N$^3$LO asymmetric matter EOS to the EOS calculated at lower order N$^2$LO. The proton drip region at N$^2$LO extends to larger proton fractions and densities but otherwise is very similar to that at N$^3$LO. This also indicates that the EFT expansion works well, as it is to be expected at these densities~\cite{Drischler2019,Keller2023}.

We also compare the results of the EFT calculations with those using a phenomenological parametrization of the energy per particle $\epsilon(n,x)$ from Ref.~\cite{Hebeler2013} [see Eq.~(2) therein], which includes the kinetic energy plus interaction terms quadratic in the neutron excess $(1-2x)$. The four parameters of $\epsilon(n,x)$ are fit to nuclear saturation, $\epsilon(n_0,1/2) = - 16\,\mathrm{MeV}$ and $P(n_0,1/2) = 0$, and by specifying the symmetry energy $S_v$ and its density derivative $L$ at saturation density. By varying $S_v$ and $L$ we can change the properties of neutron-rich matter, and study whether the proton drip region exists for reasonable ranges of $S_v$ and $L$. This is shown in Fig.~\ref{fig:robustness} for the symmetry energy $S_v = 30$ and $33\,$MeV and $L=40, 60,$ and $80\,$MeV, which represents a reasonable range based on {\it ab initio} calculations and nuclear experiments (see, e.g., Refs.~\cite{Latt12esymm,Huth2021,Drischler2021ARNPS}). Figure~\ref{fig:robustness} shows that the existence of proton drip is robust, but the exact location and extent of the proton drip phase depend on $S_v$ and $L$.

\prlsection{Analytical considerations} We now present a simple model that captures the essential features of the phase diagram. For low proton densities, the energy density of nuclear matter may be written in the form
\be
\varepsilon(n_\rn,n_\rp) \approx \varepsilon^{(0)}(n_\rn) + \mu^{(0)}_\rp n_\rp + A \, n_\rp^{5/3} + \frac12 B^{(0)} n_\rp^2 \,,
\ee
where $\varepsilon^{(0)}(n_\rn)$ is the energy density of pure neutron matter, and $\mu^{(0)}_\rp(n_\rn)$ is the proton chemical potential in pure neutron matter. The $n_\rp^{5/3}$ term comes from the kinetic energy of the protons with $A= 3 (3\pi^2)^{2/3}/(10 m_\rp^*)$ and proton effective mass $m_\rp^*(n_\rn)$. The $B^{(0)}$ term is the proton--proton interaction energy, and the superscripts~$^{(0)}$ indicate that the quantities are for pure neutron matter.
Inspired by the work of Ref.~\cite{Bardeen1967} on dilute solutions of $^3$He in superfluid $^4$He, we use the neutron chemical potential, rather than the neutron density, as an independent variable, in addition to $n_\rp$. In this way the condition for equality of the neutron chemical potentials of the two phases is satisfied automatically, thereby reducing by one the number of equations to be solved to satisfy equilibrium. To first order in $n_\rp$ the neutron chemical potential is given by
\be
\mu_\rn(n_\rn, n_\rp) = \frac{\partial \varepsilon}{\partial n_\rn} \approx \mu_\rn^{(0)}(n_\rn) + \frac{\partial\mu_\rp^{(0)}}{\partial n_\rn} n_\rp \,. 
\ee
For fixed $\mu_\rn$, on expanding $\mu_\rn^{(0)}$ to first order in $n_\rn-n_{\rn,0}$, where $n_{\rn,0}$ is the density of pure neutron matter for which $\mu_\rn^{(0)}(n_{\rn,0}) = \mu_\rn$, one finds
\be
0 = \frac{\partial \mu_\rn^{(0)}}{\partial n_\rn}(n_\rn-n_{\rn,0}) +  \frac{\partial \mu_\rp^{(0)}}{\partial n_\rn}n_\rp \,,
\ee
where the partial derivatives are to be evaluated for neutron density $n_{\rn,0}$. Thus we have
\be
n_\rn - n_{\rn,0} \approx -\frac{\partial \mu_\rp^{(0)}/\partial n_\rn}{\partial \mu_\rn^{(0)}/\partial n_\rn} n_\rp \,.
\label{deltann}
\ee

Increments in the pressure are given by $dP = n_\rn d \mu_\rn + n_\rp d\mu_\rp$. At fixed $\mu_\rn$, $dP = n_\rp d\mu_\rp$, and therefore
\be
dP = n_\rp d\left( \mu_\rp^{(0)}(n_\rn) + \frac53 A(n_\rn) \, n_\rp^{2/3} + B^{(0)}(n_\rn) \,  n_\rp\right).
\label{dp}
\ee
Expanding about the neutron density $n_{\rn,0}$, inserting Eq.~(\ref{deltann}) into Eq.~(\ref{dp}), and integrating with respect to $n_\rp$, one finds for fixed $\mu_\rn$ and to second order in $n_\rp$ that
\be
P = P^{(0)} + \frac23 A(n_\rn) \, n_\rp^{5/3} + \frac12 B \, n_\rp^2 \,,
\label{press}
\ee
where $P^{(0)}$ is the pressure of pure neutrons with chemical potential $\mu_\rn$ and
\be
B = B^{(0)} - \frac{(\partial \mu_\rp^{(0)}/\partial n_\rn)^2}{\partial \mu_\rn^{(0)}/\partial n_\rn}
\ee
is an effective proton--proton interaction when the neutron chemical potential is held constant, thereby allowing for the adjustment in the proton density due to the presence of protons. In nuclear physics terminology, $B^{(0)}$ corresponds to the ``direct'' interaction and the second term to the ``induced'' interaction due to exchange of density fluctuations~\cite{BabuBrown1973}. The discussion here parallels that for the interaction between two $^3$He atoms in superfluid liquid $^4$He~\cite{Bardeen1967}.

With the pressure given by the form (\ref{press}), it is impossible to satisfy the conditions for phase equilibrium: if $B$ is positive, the pressure increases monotonically with $n_\rp$, so the condition for pressure equilibrium cannot be satisfied, while if $B$ is negative, there are two values of $n_\rp$ for which the pressures are equal, but the state with higher $n_\rp$ is unstable, in that $\partial P/\partial n_\rp|_{\mu_\rn}$ is negative and the system will collapse to high proton densities.
To satisfy the conditions for phase equilibrium, it is necessary to have an additional positive contribution to the pressure that increases with $n_\rp$ faster than $n_\rp^2$. For definiteness, we add a contribution $2 C n_\rp^3$ to the pressure (corresponding to a term $C n_\rp^3$ in the energy density), although the qualitative form of the phase diagram does not depend on the choice of the power.
The first term in Eq.~(\ref{press}) plays no role in the conditions for coexistence of two phases, and we shall drop it.  On introducing dimensionless variables, we may write the pressure due to the protons as
\be
P_\rp (y, \nu) = \frac23 A \tilde n^{5/3} (y^{5/3} - 2 \nu y^2 + y^3) \,,
\ee
where $y=n_\rp/\tilde n$ with $\tilde n=(A/3C)^{3/4}$, and $\nu =- 3^{3/4} B/(8 A^{3/4} C^{1/4})$. The proton chemical potential $\tilde{\mu}_\rp$ relative to $\mu_\rp^{(0)}(n_\rn)$ is given from Eq.~(\ref{dp}) by
\be
\tilde{\mu}_\rp (y, \nu) = \frac23 A \tilde n^{2/3} \left(\frac52 y^{2/3} - 4 \nu y + \frac32 y^2\right).
\ee
Thus a single dimensionless parameter $\nu$ governs the phase diagram.  

\begin{figure}
  \centering
  \includegraphics[clip=,width=\linewidth]{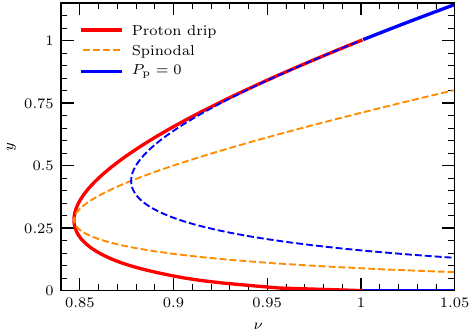}
  \caption{Proton densities $y$ (in units of $\tilde n$) of phases in equilibrium as a function of $\nu$ in our simple model. The solid red line shows the two nonzero proton densities for which coexistence (proton drip) is possible. The blue line shows the values of the proton density for which pressure equilibrium of nuclear matter and pure neutron matter (neutron drip) is possible ($P_\rp=0$); the solid line indicates where the equilibrium is stable and the dashed line where it is unstable. The dashed orange curve is the spinodal line, $\partial P_\rp/\partial n_\rp|_{\mu_\rn}=0$, where uniform matter would become unstable.}
  \label{fig:analytical-model}
\end{figure}

Let us consider how the proton density varies with $\nu$ starting from large values. The condition for equilibrium of nuclear matter with pure neutrons is that the pressure due to the protons is zero, or $y^{5/3} - 2 \nu y^2 + y^3=0$. The lowest value of $\nu$ for which this equation has a solution is $\nu=2/3^{3/4} \approx 0.8774$. However, for $\nu=1$ (when $y=1$) the proton chemical potential in nuclear matter becomes equal to that in the pure neutron phase $\mu_\rp^{(0)}(n_\rn)$ (or $\tilde{\mu}_\rp=0$), and for lower values of $\nu$ the two phases in equilibrium both contain protons. The proton densities in the two phases, $\tilde n y_1$ and $\tilde n y_2$, are given from the equality of pressures and proton chemical potentials:
\be
P_\rp(y_1, \nu) = P_\rp(y_2, \nu) \quad \text{and} \quad
\tilde{\mu}_\rp (y_1, \nu) = \tilde{\mu}_\rp (y_2, \nu) \,.
\ee
The lowest value of $\nu$ for which a real solution is possible is $5^{3/4}/3^{5/4}\approx 0.8469$ (this is also the same for which the spinodal instability condition $\partial P_\rp/\partial n_\rp|_{\mu_\rn}=0$ has a solution). The results of our calculations are shown graphically in Fig.~\ref{fig:analytical-model}. In summary, stable equilibrium is possible between nuclear matter and pure neutron matter for $\nu>1$ and between two phases both of which have nonzero proton concentrations for $5^{3/4}/3^{5/4} < \nu < 1$. For $\nu < 5^{3/4}/3^{5/4}$ matter consists of a single phase. We have checked that fits of $A$, $B$, and $C$ to our asymmetric matter calculations at densities around $n_0/2$ lie within the range where proton drip occurs in the analytical model. Therefore, we conclude that our simple model also exhibits a stable proton drip phase.

\prlsection{Discussion and conclusions} In this Letter we have used state-of-the-art chiral EFT calculations of asymmetric nuclear matter to explore the phase diagram for neutron-rich conditions at zero temperature, relevant for cold neutron stars. This automatically takes into account the non-analytical features of the EOS at low proton concentrations, unlike essentially all earlier works, which were based on a simple interpolation between the properties of symmetric matter and pure neutron matter.

A striking finding of our work is the occurrence of proton drip, the coexistence of two phases of bulk nuclear matter with different nonzero concentrations of protons. We have shown that this is a general feature of a variety of EOSs, including ones based on interpolation between symmetric nuclear matter and pure neutron matter. At nonzero temperature, the low-density matter in the proton drip phase would also include clusters (deuterons, $^3$H, $^{3}$He, $^{4}$He)~\cite{Virial2006,Arcones2008}. This will be an interesting topic for future explorations. However, proton drip occurs over a limited range of conditions: In Ref.~\cite{Pethick1995}, the uppermost pressure for which proton drip occurred was only 7\% greater than the lowest pressure for proton drip. We thus attribute the fact that proton drip was not found in Ref.~\cite{BBP} to the pressures for which it occurs lying between those of the grid points used.

We have also shown how proton drip can be understood on the basis of a simple analytical model. The starting point of the model is the energy of the uniform phase as a function of the proton density at constant neutron chemical potential. The basic ingredients are the proton kinetic energy, an effective two-body attraction between protons, and a repulsive contribution to the energy varying as the proton density to a power higher to ensure that the system does not collapse to high proton density.

Armed with the new results for the EOS, it is now possible to make a renewed attack on the problem of whether or not phases with string-like and plate-like nuclei (pasta phases) are stable in neutron stars. Pasta phases in neutron stars could have a significant effect on observable properties. Pasta phases are stable at higher proton concentrations, but for the low proton concentrations in neutron star matter in beta equilibrium, the energy difference between the uniform phase and the two-phase state is much smaller: it could be less than the Coulomb and surface energy cost of making the two-phase state, as was found in Ref.~\cite{Douchin2001}.

We have performed preliminary calculations of the filling factor $u$ for neutron star matter in the phase of neutron and proton drip, i.e., along the beta-equilibrium line in Fig.~\ref{fig:phase-diagram} neglecting surface and Coulomb effects. We find that the filling factor increases with increasing density toward the boundary of the neutron drip phase to $u \approx 0.2$. In the proton drip phase it increases further to maximum values around $u \approx 0.4$. The larger the filling factor, the more likely pasta phases are, so that proton drip aids the existence of pasta phases (see, e.g., Fig.~1 of Ref.~\cite{Ravenhall1983}, where $n/n_s$ corresponds to $u$). An important future task is thus to evaluate the nuclear surface energy for matter with low proton concentrations consistently in chiral EFT. Moreover, it would be interesting to explore magnetic properties in the neutron star crust for which superfluid protons in the proton drip phase can play an important role.

This work was supported in part by the Deutsche  Forschungsgemeinschaft  (DFG, German Research Foundation) -- Project-ID 279384907 -- SFB 1245 and the European Research Council (ERC) under the European Union's Horizon 2020 research and innovation programme (Grant Agreement No.~101020842).  NORDITA is supported in part by NordForsk.

\bibliography{literature}

\end{document}